# Ultra-Broadband Terahertz Perfect Absorber based on Doped Silicon

Ankit Vora
Dept. of Elec. & Comp. Eng.
Michigan Tech. Univ.
Houghton, MI, USA
avora@mtu.edu

Satyadhar Joshi
Grad. School of Tech.
Touro College
New York, NY, USA

Arun Matai
Dept. of Elec. & Comm. Eng.
Acropolis Tech. Campus
Indore, MP, INDIA

*Abstract*— The requirement for metamaterial perfect absorbers (MPA) based on doped semiconductors is steadily increasing due to the available matured fabrication and simulation technology. There is a particular interest in developing terahertz (THz) perfect absorbers using doped semiconductors for achieving characteristics such as polarization-independence, wide-angle, and broadband absorption. We report MPA based on patterned arrays of tapered micro-cylindrical structures of doped silicon to enable them with broadband, wide-angle, and polarization-independent response. In this work, we modeled the MPA structures using COMSOL Multiphysics® to evaluate its electromagnetic wave response using the software's RF Module running in parallel on a Beowulf cluster. We evaluated the doped silicon MPA structures for its response in the frequency spectrum of 0.1 to 5.0 THz for transverse magnetic and transverse electric polarizations at the normal and oblique incidence up to 75 degrees. This proposed doped silicon MPA was found to support a perfect absorption for a wide frequency spectrum from 1.7 to 3.9 THz along with insensitivity towards polarization and incident angles up to 60°. The execution of MPA on simplified Beowulf cluster significantly reduced the simulations time by the orders of magnitude compared to the sequential simulations.

## I. Introduction

In the past decade, metamaterials have been explored to a great extent due to their phenomenal ability to manipulate light. Metamaterials are judiciously engineered structures of optical materials capable of responding to any region of the electromagnetic spectrum and support optical properties such as cloaking, polarization-independence, wide-angle, and broadband optical response for a diverse list of applications [1-2]. The necessity for metamaterial perfect absorbers operating in the terahertz (THz) part of the frequency spectrum is increasing for cloaking, sensors, and imaging applications. Silicon is becoming a preferred candidate for THz based perfect absorbers being able to respond to a wide spectrum of electromagnetic radiation by varying the doping concentration and available matured fabrication technology [3]. Doped silicon has proven to exhibit surface plasmon polaritons and subsequently localized surface plasmon resonances through periodic structures and consequently behaves as an excellent lossy material in the THz range. Microfabrication of silicon-based devices is ubiquitous and extensively explored. Thus, the infrastructure present at both the simulation and fabrication level makes designing THz material with desired electromagnetic property convenient [4].

As the most elementary design, one-dimensional (1-D) absorbers are simple to fabricate but their response is typically polarization and incident-angle dependent, whereas two-dimensional designs are more promising when it comes to performance, but they are still sensitive to the polarization and the incident angle of the electromagnetic radiation. On the other hand, three-dimensional (3-D) absorbers can not only provide a true polarization and incident-angle independent response, but they are also consistent with current microfabrication technology. In our previous work [5-12], we extensively studied patterned arrays of silver nano-cylinders for improving the conversion efficiency of hydrogenated amorphous silicon solar cells. These nano cylinders endowed the solar cells with multi-resonant, polarization-independent, and wide-angle response. This concept can also be extended to THz absorbers using patterned micro-cylinders which can aid in superior broadband absorption characteristics. Furthermore, including tapered cylinder design can drastically improve the absorption characteristics.

In this work, we have explored 3-D absorbers based on doped silicon MPAs using periodic arrays of tapered cylindrical structures [13]. Absorption inside the doped silicon MPA structures can be tuned for a frequency spectrum by varying the doping concentration and changing the physical parameters of the MPA structures. However, due to the complex design of the proposed structures, sequential evaluation and optimization becomes challenging both temporally as well as computationally. Therefore, in order to estimate the performance of the proposed doped silicon MPA structures, we implemented the simulation model on a Beowulf cluster running COMSOL Multiphysics in parallel. In this work, we have demonstrated the superior performance of the doped silicon MPA structures portraying wideband, wide-angle, and polarization insensitive response of the proposed design.

## II. METHODS

The complex permittivity of the doped silicon is portrayed by the Drude model [14] using the equation:

$$\varepsilon = \varepsilon_\infty - \frac{\omega_p^2}{\omega^2 + i\gamma\omega} \quad (1)$$

where $\varepsilon_\infty = 11.7$ is the free space dielectric constant of the intrinsic silicon, $\omega_p$ is the plasma frequency, and $\gamma$ is the carrier scattering rate. The plasma frequency and the carrier scattering rate are a function of carrier concentration and mobility of the doped silicon, which can be experimentally measured through Hall effect analysis. In this paper, we choose $\omega_p = 2\pi \times 5.22$ THz and $\gamma = 2\pi \times 1.32$ THz for doped silicon MPA modeling as reported in [14].

This particular geometry of tapered cylindrical structures was selected due to its top structure axial isotropy that leads to a polarization independent and broadband optical response [5]. The structure of our doped silicon MPA are schematically shown in Fig. 1(a), and the unit cell of this doped silicon MPA is shown in Fig. 1(b). The entire structure was modeled using heavily doped silicon. In this paper, the doped silicon MPA is modeled using tapered cylinders of the height ($h_c$) 140 μm, the bottom width ($r_b$) of 90 μm, the top width ($r_t$) of 40 μm, and periodically spaced with period (p) 100 μm. The thickness of the substrate ($h_b$) under the tapered cylindrical structures is fixed at 200 μm which are large enough to eliminate transmission losses.

Optical responses of the doped silicon-based MPA structures were obtained by modeling its unit cell in RF module of COMSOL Multiphysics v5.3b in the frequency domain. Floquet boundary conditions were employed for the side faces of the unit cell. For the top face, excitation port was deployed with diffraction to include the influence of the scattered waves, and for the bottom face, an output port was employed. The excitation port and the output port were suspended in a thin column of air in the model. The $S_{11}$ and $S_{21}$ parameters were employed to calculate the frequency dependent absorbance, $A(\omega)$, transmittance, $T(\omega)$, and reflectance, $R(\omega)$, for the transverse magnetic (TM) and the transverse electric (TE) incident waves.

$$\begin{aligned} T(\omega) &= |S_{21}(\omega)|^2 \\ R(\omega) &= |S_{11}(\omega)|^2 \\ A(\omega) &= 1 - T(\omega) - R(\omega) \end{aligned} \quad (2)$$

The model was simulated for a frequency spectrum of 0.1 to 5.0 THz and an oblique incidence of 0 to 75°. For qualitative analysis, the power dissipated in the entire structure was also studied using the inbuilt function power loss density in COMSOL Multiphysics RF module.

COMSOL was executed parallelly using distributed-memory parallel operations. The simulations were executed on 2 Intel Xeon E5 processors with 16 cores running at 2.50 GHz and 64 GB of RAM for each processor. COMSOL has an inbuilt support for message passing interface (MPI) which enables distributed computations in parallel. Hence, COMSOL was set up to run on a Beowulf cluster with COMSOL's inbuilt cluster computing features.

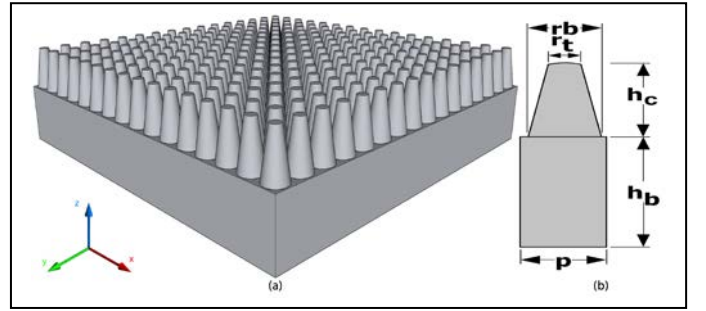

Fig 1. (a) The MPA structures based on doped silicon (b) The unit cell of doped silicon MPA structures with geometric parameters.

## III. RESULTS AND DISCUSSION

The optical response of the doped silicon MPA for normal and oblique incidence is shown in Fig. 2, and Fig. 3 respectively. For comparison we also simulated a bare silicon substrate having identical optical properties and thickness (340 μm). Near perfect absorption has been observed for a broad frequency spectrum ranging from 1.7 to 3.9 THz for our proposed doped silicon MPA structures. This optical absorption was found to be uniform for both normal as well as oblique angles of incidence. Besides that, MPA structures exhibited insensitivity towards polarization and incidence angles up to 60° for both TM and TE polarizations. This bare silicon substrate is strongly absorbing in the 1.8 to 4.9 THz region of the spectrum; however, the maximum absorbance in the case of the bare silicon substrate is 76% compared to almost perfect absorbance for the doped silicon MPA structures. It is evident from the Fig. 2 that the doped silicon MPA outperforms the bare silicon substrate for all parts of the spectrum, especially in the region of 1.0 to 4.5 THz where absorbance is greater than 95% because of the broad multi-resonant response.

The optical response of doped silicon MPA has several resonant peaks which need to be examined to understand the physical mechanism responsible for such broadband absorption. We analyzed the magnetic field distribution (Fig. 4) at all the peaks in optical response of the doped silicon MPA, i.e., at resonant frequencies: 1.0 THz, 1.8 THz, 2.7 THz, and 3.7 THz. The underlying mechanism for near unity broadband perfect absorption was found to be a combination of Fabry-Perot resonance, mode-matching resonance, and air-cavity mode based on the analysis of field distribution and power density inside MPA structures.

For the low-frequency peaks, the incident magnetic field is mainly concentrated in the air cavities formed between the adjacent silicon tapered cylinders as evident from Fig. 4 (a) and (b). We also calculated the power loss density inside the doped MPA structures and from Fig 4 (e) and (f) it can be further confirmed that the loss is concentrated at the bottom corner of silicon groove since the incident waves are guided to the substrate. Also, for the absorption peaks at 1.0 and 1.8 THz, the incident wave is mode-matched with the top layer of silicon, and hence most of the electromagnetic power is consumed by the top layer of silicon as evident from Fig. 4 (e) and (f). Therefore, the peaks at these resonant frequencies can be attributed to air cavity mode and mode matching resonances. The doped silicon metasurfaces also provide impedance matching condition to the free space for both TM and TE polarizations and thus enabling efficient coupling and the high absorption coefficient of silicon also assist in improving the absorption along the direction of wave propagation. We also analyzed the magnetic field distribution at resonant frequencies 2.7 and 3.7 THz, Fig. 4 (c) and (d), and this analysis reveals the presence of Fabry-Perot like resonance inside the doped MPA structures which is similar to the field distribution as reported in [5] and [15].

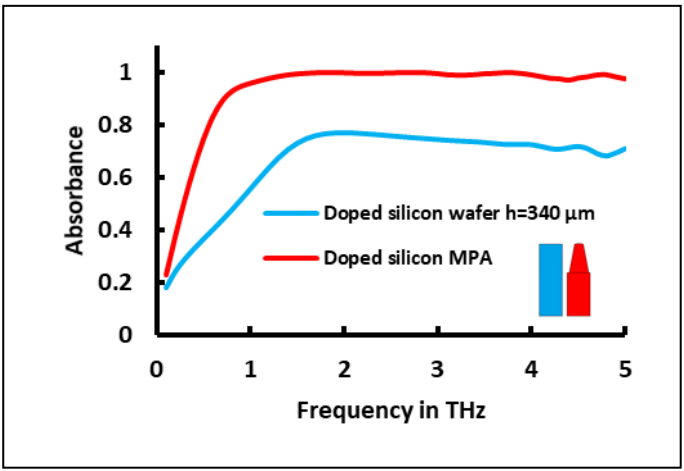

Fig. 2 Absorbance in doped silicon MPA and doped silicon wafer for normal incidence

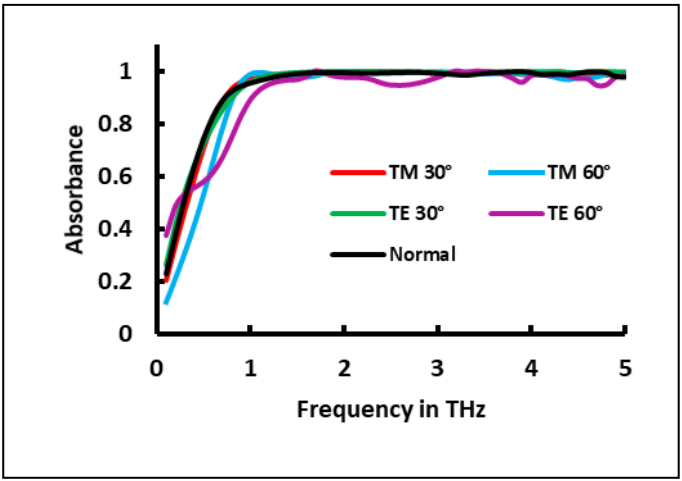

Fig. 3 Absorbance in doped silicon MPA for oblique incident waves for TE and TM polarizations

The proposed doped silicon MPA structures are not only superior in performance compared to the bare silicon substrate but also have an advantage of ease of fabrication and scaling with the current microfabrication technology. Secondly, this structure can be fabricated in a very few steps and consequently decrease the overall fabrication cost. This design can be integrated with applications ranging from sensors, THz imaging, perfect-absorbers, biomedical, and cloaking.

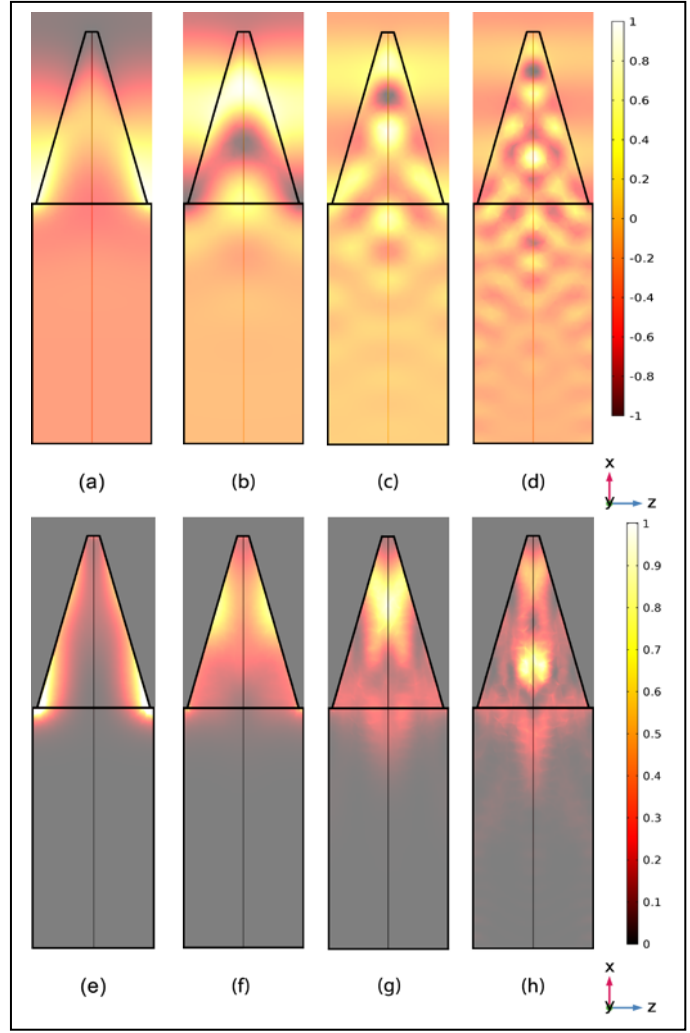

Fig. 4 The field distribution in the doped silicon MPA structures (magnetic, $H_y$) (a-d), and the power density in the doped silicon MPA structures (e-h) at resonant frequencies: {1.0, 1.8, 2.7, 3.7} THz

## IV. CONCLUSIONS

In summary, we discussed the MPAs based on patterned arrays of tapered cylindrical structures of heavily doped silicon. This doped silicon MPA supports characteristics such as wideband, wide-angle, and polarization insensitive response suitable for applications such as sensors, imaging, radar cloaking, etc. The design of this structure is within the permissible limits of the current microfabrication technology. This terahertz absorber supports near-perfect absorption for a broad spectrum of 1.7 to 3.9 THz for a wide incidence angle up to 60°. This doped MPA structure was modeled in COMSOL Multiphysics in RF module running on a Beowulf cluster for reducing the overall simulation time.

## ACKNOWLEDGMENT

We thank Jay Jagannath Das and Himanshu Kothari for insightful discussions.